\documentclass[acmsmall,nonacm,anonymous=false]{acmart}

\usepackage{amsmath}
\usepackage{array}
\usepackage{enumitem}
\usepackage{multirow}
\usepackage{subcaption}
\usepackage{tikz}
\usepackage{xfrac}

\setlist[enumerate]{leftmargin=*}
\setlist[itemize]{leftmargin=*}

\graphicspath{{./}{figures/}}

\begin{document}

\title[Scoring the Unscorables: Cyber Risk Assessment Beyond Internet Scans]{Scoring the Unscorables:\\Cyber Risk Assessment Beyond Internet Scans}

\author{Armin Sarabi}
\email{arsarabi@umich.edu}
\affiliation{%
  \institution{University of Michigan}
  \city{Ann Arbor}
  \state{MI}
  \country{USA}
}

\author{Manish Karir}
\email{mkarir@signetrisk.com}
\affiliation{%
  \institution{SignetRisk Analytics, Inc.}
  \city{Ann Arbor}
  \state{MI}
  \country{USA}
}

\author{Mingyan Liu}
\email{mingyan@umich.edu}
\affiliation{%
  \institution{University of Michigan}
  \city{Ann Arbor}
  \state{MI}
  \country{USA}
}

\begin{abstract}
In this paper we present a study on using novel data types to perform cyber risk quantification by estimating the likelihood of a data breach. We demonstrate that it is feasible to build a highly accurate cyber risk assessment model using public and readily available technology signatures obtained from crawling an organization's website. This approach overcomes the limitations of previous similar approaches that relied on large-scale IP address based scanning data, which suffers from incomplete/missing IP address mappings as well as the lack of such data for large numbers of small and medium-sized organizations (SMEs). In comparison to scan data, technology digital signature data is more readily available for millions of SMEs. Our study shows that there is a strong relationship between these technology signatures and an organization's cybersecurity posture. In cross-validating our model using different cyber incident datasets, we also highlight the key differences between ransomware attack victims and the larger population of cyber incident and data breach victims.
\end{abstract}


\maketitle

\section{Introduction}\label{sec:intro}
The ability of machine learning (ML) algorithms in analyzing large amounts of data and identifying patterns enables researchers to uncover correlations that can estimate and forecast risks with substantial accuracy. The resulting data-driven insights allow for more effective and informed risk management strategies by adapting to the ever-evolving threat landscape. Prior work in this direction include estimating organizations' likelihood of experiencing a material data breach~\cite{liu2015cloudy,sarabi2016risky,kure2022asset}, estimating the likelihood of software vulnerability (CVEs) exploitation~\cite{sabottke2015vulnerability,tavabi2018darkembed,jacobs2021exploit,suciu2022expected,jacobs2023enhancing}, predicting security events (e.g., infection) on individual machines~\cite{bilge2017riskteller,shen2018tiresias}, maliciousness of apps and executables~\cite{pan2017dark,rhode2018early}, and the risk of a benign website becoming malicious in the future~\cite{soska2014automatically}.

Of particular relevance to the present study is \cite{liu2015cloudy}, which demonstrates the viability of using Internet scan data to perform highly accurate data breach predictions. Internet scanning~\cite{durumeric2013zmap,durumeric2015search} has long been the bedrock of understanding vulnerabilities, attack surfaces, and overall cyber risks, both at a device level and at a network/system level. These measurements have been widely used for a variety of purposes, including to detect and fingerprint networked devices~\cite{bano2018scanning,demarinis2019scanning,feng2018acquisitional,scheitle2018first,sarabi2023llm}, study trends~\cite{felt2017measuring,kotzias2018coming,kumar2018tracking}, examine security events~\cite{antonakakis2017understanding,durumeric2014matter}, and enable various machine learning aided cybersecurity analysis \cite{liu2015cloudy,sarabi2018characterizing}. The data used in \cite{liu2015cloudy} includes Internet scans and IP address-based reputation blacklists (RBLs), such as those maintained by PhishTank~\cite{phishtank} and Spamhaus~\cite{spamhaus}, and breach reports, such as the VERIS Community Database (VCDB)~\cite{vcdb}, as labels to enable supervised learning; the prediction output is in the form of an estimated probability of an organization suffering a material data breach. This type of risk quantification marked a concrete progress towards generating a crucial type of cybersecurity actuarial data that was missing for practices such as vendor management and cyber insurance underwriting.

On the other hand, studies such as  \cite{liu2015cloudy}  heavily rely on Internet scan data.  This results in a number of limitations. The first has to do with the fact that Internet-wide scans consist of raw information obtained from protocol handshakes (e.g., banner grabs) with low label/feature coverage. This in turn has severely limited how much of the raw data makes its way into ML models. For instance, while scan data returns HTTPS banners containing a wealth of information such as server version, software and security settings, etc., the only information used in the study performed by \cite{liu2015cloudy} is the binary feature of whether there is a valid HTTPS certificate.

The second limitation is that many organizations, especially small and medium-sized enterprises (SMEs), do not have dedicated network assets, and instead utilize cloud providers such as Amazon Web Services (AWS), Microsoft Azure, and Google Cloud Platform (GCP) to host their services. Thus, the singular reliance on Internet scan data cannot afford them meaningful risk assessment.

Last but not least, even when Internet scan data is available, it is at the IP address/host level. Since data breaches are reported at the higher, organization level, scan data has to be in aggregated form so one can meaningfully match the scan data (features) with breach information (labels) to facilitate supervised learning. This aggregation involves two steps \cite{liu2015cloudy}: network asset (ownership) discovery along organizational boundaries (mapping/attributing individual IP addresses to the organizations that they belong to), and aggregate (heuristic) feature extraction using key host data elements (e.g., counting the total number of untrusted HTTPS certificates within an organization). It turns out that asset discovery and IP address attribution was and remains a particularly difficult, labor-intensive, and error-prone process, since IP address ownership can be fluid (with the sale, merger, and acquisition of assets) and obscure (transactions under different entity names and complex subsidiary relationships).

In this paper, we show how we can get around these limitations by focusing on a novel set of web crawl data. As they are directly associated with Internet domains, these data can be readily matched with breach reports to enable supervised learning exercises. We will show that this type of alternative data can produce highly accurate risk estimate results. We discuss in detail the robustness and feature importance of the trained models, and also shed light on the difference between broadly defined cyber incidents and more specific ransomware incidents from the perspective of model training.

There is an interesting analog between our study and a recent study on using grocery shopping behavior to generate consumer credit scores that are more accurate than using traditional methods \cite{lee2025using}: about 45 million US adults do not have sufficient credit history to be given a credit score, which can severely limit their ability to access the financial system. Similarly, there are over 30 million SMEs in the US~\cite{sme}, a substantial fraction of which do not have a physical Internet presence to be afforded meaningful cyber risk assessment that relies heavily on Internet scan data.

Our main contributions are summarized as follows.

\begin{enumerate}
    \item We show how a novel type of web crawl data can be used to train highly accurate cyber risk assessment (or scoring) models, thereby circumventing the major limitations of existing approaches that rely on Internet scan data.
    \item We show a high yield ($>$95\%) on collecting this type of crawl data, which means this risk assessment model can be applied to the millions of small entities that do not have a physical Internet presence.
\end{enumerate}

The remainder of the paper is organized as follows. In \autoref{sec:method} we give an overview of the data collection and model training methodology, and then detail the data collection and preparation process in \autoref{sec:data}. \autoref{sec:result} describes the training process and examines the performance,  robustness, and cross-validation of the trained models. \autoref{sec:enhancement} shows how the model performance can be further enhanced by adding auxiliary data/features. \autoref{sec:discussion} discusses additional interpretations of the model output, feature importance, and model training using historical data, and \autoref{sec:conclusion} concludes the paper.

\section{Methodology}\label{sec:method}
In this study we leverage \emph{supervised learning} approach, where features and (binary) labels are paired and associated with a given Internet domain; these feature-label pairs are then used to train a classifier. Inference for domains with known features and unknown labels results in model outputs that indicate an estimated probability of a domain being associated with a positive label.

\begin{figure}
    \centering
    \includegraphics[width=0.55\textwidth]{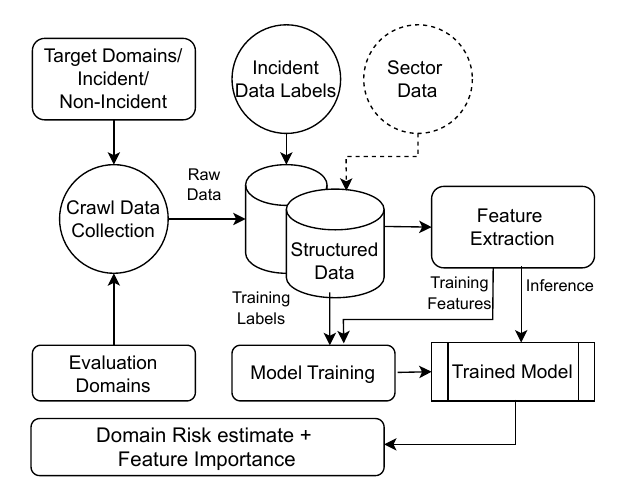}
    \caption{A domain based cyber risk prediction/assessment system. Dashed circle indicates optional data element outside the main methodology.}
    \label{fig:domain-risk}
\end{figure}

We use reports of cybersecurity incidents to associate organization domains and the occurrence of such incidents as positive labels. We then collect data from the website hosted at a domain using a web crawl agent. This is coupled with an extensive data preparation phase in order to generate the structured features needed to train our classifier. Both types of data are described in detail in the next section. \autoref{fig:domain-risk} gives a high-level view of our overall methodology.

\subsection{Modeling at the domain level} The data used for building a model is derived from a variety of different sources and it is important to normalize them to ensure that the different characteristics being represented are referencing the same entity.  We use domain names as the canonical identifier for this purpose. The feature data we collect is naturally aligned with specific domain names, and any victim entity names identified in cyber incident reports are also mapped to the associated domain for that entity. Unlike organization names, domain names are guaranteed to be globally unique. This ensures a strong consistency in our feature and label datasets.

Using domain names also eliminates significant challenges with other approaches as mentioned in \autoref{sec:intro}. One such prior approach~\cite{liu2015cloudy} attempts to create mappings between entity names and their associated IP address ranges, creating a number of technical challenges as described below.

\begin{itemize}
    \item There is no single global directory of IP address assignment and ownership.  There is a system of Regional Internet Registries (RIRs), one being the American Registry for Internet Numbers (ARIN) for North America, that record IP address allocations at a high level. However, with the extensive use of business Internet connections and intermediate Internet Service Providers (ISPs), these databases stop at ISP address delegations, and most smaller organizations are simply not represented. Additionally, even for larger organizations, there is no mandatory requirement for them to keep their IP address ownership information up to date, leading to information degradation over time.
    \item Most modern organizations utilize large scale infrastructure/platform as a service (IaaS/PaaS) providers for at least some if not most of their infrastructure, these platforms often rapidly rotate IP addresses as part of their normal service operation.  This makes it even more challenging to create accurate mappings between a given entity and their IP address footprint.
\end{itemize}

Also, as mentioned earlier, data breaches and other types of cybersecurity incidents are almost always associated with an organization, not a specific IP address, making it difficult to align model features with incident labels. The result of these challenges is that the applicability of cyber risk assessment based on this data is limited to a small set of organizations where such IP address mappings can be adequately determined.  This leads to a situation where millions of organizations cannot be evaluated due to the lack of availability of IP address data. This is a limitation that has persisted over a long time, and is a strong motivation for us to explore an alternate path based on domain names.

By using domain names for both feature and label information, we avoid the attribution problem, and the association between the two becomes much more straightforward.  This approach also opens up entirely new categories of domain-level data that can be used for risk quantification.  Moreover, models built on such features are more robust and broadly applicable to a broader range of organizations as we discuss in \autoref{sec:performance}.

\subsection{Unlocking new data and model features}

By focusing on data and features that are available at the domain level, we are able to tap into a rich set of novel data. Using a web crawler, we are able to uncover a large number of technological characteristics of an organization's website. This data provides us with a new lens into a wide range of cybersecurity-related practices. Below we briefly list some typical elements. A detailed list is provided in Table 1.

\begin{itemize}
    \item Security technologies in the form of botnet traffic mitigation, cookie management, and fraud prevention, such as Cloudflare bot management, AWS WAF CAPTCHA, CookieYes, or ThreatMetrix.
    \item Software stack and library dependencies in the form of content management systems, JavasScript libraries, and databases, such as Drupal, Joomla, Lodash, Vue, or MySQL.
    \item The use of analytics/pixel trackers in the form of web analytics, SEO, or Real User Monitoring (RUM) technologies, such as AppDynamics, DynaTrace, Facebook Pixel, Google Tag Manager, or Adobe Analytics
    \item Internet hosting platforms in the form of CDNs, load balancers, and hosting platforms, such as Cloudflare, Fastly, F5, Nginx, or WPEngine.
    \item Financial elements in the form of advertising, payment, and shopping cart related technologies, such Google AdSense, AdScale, Klarna, Moneris, or Mulberry .
    \item Client engagement and support technologies in the form of marketing automation, comment systems, and issue trackers, such as Sentry, HubSpot Analytics, or Constant Contact.
\end{itemize}

In each instance the presence or absence of a technology (also referred to as a technology signature interchangeably) can shed light on the level of cybersecurity sophistication at an organization.  Some of these are considered common best practices (e.g., Google Tag Manager), some provide advanced cybersecurity protections (e.g., Cloudflare), while some are frequently associated with recurring vulnerabilities (e.g., WordPress). Our goal is to see whether a wide collection of such data can generate meaningful risk assessment.

\section{Data Collection and Preparation}\label{sec:data}
In this section we detail our data acquisition process. After describing how we collect the feature and label data, we also provide statistics on the yield of our crawl method, i.e., the fraction of samples that our methodology can return valid information that can be used for model training and evaluation.

\subsection{Raw feature data: technologies}\label{sec:crawl}

Our (raw) feature data is collected through a web crawl from a given domain/URL; we explain in detail how these domains are chosen in \autoref{section:labels}. The web crawl is implemented based on a number of commonly used techniques, and relies on a crawler agent to visit an organization's website (URL) and automatically navigate to a number of random and targeted subpages. During this process, the crawler agent is designed to identify and record a set of digital signatures that uniquely identify specific technologies that are being used on that website. These technologies primarily consist of externally observable components, such as those exposed through HTTP headers, HTML content, embedded JavaScript, and so on. The types of technologies that we identify were outlined in \autoref{sec:method}, and we will detail specific examples later in this section through a specific case study.

Specifically, in order to implement a scalable and flexible data collection architecture, we use the Crawlee library in conjunction with Playwright and headless Chromium browser as the back-end.\footnote{\url{https://crawlee.dev}, \url{https://playwright.dev}} This allows us to accurately navigate websites that use JavaScript in an automated manner.  Digital signature extraction is implemented with the help of the Wappalyzer Chrome extension.\footnote{\url{https://github.com/dochne/wappalyzer}} Data collection starts with the crawler agent visiting the homepage of a given organization's website. Additional pages to be crawled are identified and added to the crawler queue according to the following strategies/heuristics.

\begin{itemize}
    \item \textbf{Random navigation}: We identify embedded references to pages within the same domain and add \textbf{9} such links (with a maximum of 3 links extracted from a single web page) at random to the list of follow-on pages that the crawler agent will visit. This random navigation to additional internal pages helps to balance the visibility of the crawler  to a large number of potential web pages with the need to minimize overhead traffic to the target website.  The range of pages visited by the crawler is intended to increase the accuracy of technology detection in cases where a technology is utilized on a page other than the website's landing page.
    \item \textbf{Targeting privacy pages}: In addition to the above strategy, our crawler agent also identifies up to \textbf{9} internal links containing the word ``privacy''. Other variations that the crawler agent attempts to visit include \texttt{/privacy-policy} and \texttt{/privacy} subpages. Here, the presence of a privacy policy is used as a heuristic/proxy for identifying websites associated with an organization versus those that do not (more on this in \autoref{section:negatives}).
\end{itemize}

\newcommand\Tstrut{\rule{0pt}{2ex}}
\newcommand\Bstrut{\rule[-1ex]{0pt}{0pt}}

\begin{table*}[t]
    \small
    \centering
    \caption{List of technology categories and meta-categories. These technology signatures are obtained using the crawl process described in \autoref{sec:crawl}. Numbers in parentheses correspond to the number of features extracted from each category/meta-category using the extraction process described in \autoref{sec:extract}. We prune our feature set to only include those that are observed (i.e., non-zero) for at least 20 of our crawled websites.}
    \label{table:technologies}
    \begin{tabular}{|l|p{8.6cm}|}
        \hline
        \multicolumn{1}{|c|}{\bf Meta-category} &
        \multicolumn{1}{c|}{\bf Categories} \\
        \hline
        Software Stack (630) & JavaScript libraries (180), WordPress plugins (108), JavaScript frameworks (92), UI frameworks (38), Web servers (34), CMS (33), Font scripts (27), Programming languages (23), Widgets (22), Web frameworks (21), Form builders (17), Video players (15), WordPress themes (15), Blogs (12), Page builders (12), Static site generator (11), Development (10), Caching (9), CRM (8), Web server extensions (8), Databases (7), Operating systems (7), Maps (6) \\
        \hline
        Web Analytics/Pixel Trackers (115) & Analytics (75), SEO (25), RUM (11), Tag managers (6) \\
        \hline
        Miscellaneous (101) & Miscellaneous (23), JavaScript graphics (9), Editors (8), Photo galleries (7), Translation (7), Accessibility (5), Reviews (5), Rich text editors (5), Appointment scheduling (4), Digital asset management (4), Mobile frameworks (3), Search engines (3), Browser fingerprinting (2), Documentation (2), Fundraising \& donations (2), Hosting panels (2), Recruitment \& staffing (2), User onboarding (2), Content curation (1), Geolocation (1), Loyalty \& rewards (1), Message boards (1), Shopify apps (1) \\
        \hline
        Financial Elements (82) & Advertising (41), Ecommerce (14), Payment processors (13), Retargeting (6), Affiliate programs (5), Shipping carriers (5), Buy now pay later (2), Cart abandonment (1) \\
        \hline
        Customer Support (75) & Marketing automation (21), Live chat (13), A/B Testing (12), Personalisation (11), Customer data platform (8), Issue trackers (7), Comment systems (5), Segmentation (4), Surveys (4) \\
        \hline
        Internet Hosting (67) & CDN (18), Performance (15), PaaS (12), Reverse proxies (10), Hosting (7), IaaS (3), Load balancers (3), Livestreaming (2) \\
        \hline
        Security/Privacy (37) & Cookie compliance (20), Security (11), Authentication (5) \\
        \hline
        Communication Systems (6) & Email (5) \\
        \hline
    \end{tabular}
\end{table*}

The identified technologies through the above crawl process are then mapped to a set of categories by Wappalyzer. This is listed in \autoref{table:technologies} (right column), which we further organize into 8 broad, meta-categories (left column).  The organization or grouping of categories into meta-categories serves two purposes. First, it allows us to add additional numerical features as detailed in \autoref{sec:extract}, aimed at helping the algorithm learn to recognize the amount of different technologies associated with different aspects of cyber risk.  Secondly, it allows us to better interpret the classifier output when we examine feature importances in \autoref{sec:performance}.

\paragraph{How technology signatures are collected}: We step through a specific case to illustrate how the crawler works and provide an intuitive understanding of why we believe this type of data holds value for estimating cyber risk. Consider the example of ``23andMe Holding Co'', a relatively modern publicly traded company with over 500 employees and annual revenues of several hundred million dollars. A quick analysis of their website using our crawler agent reveals the following:

\begin{itemize}
    \item They use 9 different types of analytics/pixel trackers of website visitors such as Facebook, Google, Auryc, and Adobe.  These are often easily identified by specific JavaScript files or signatures.
    \item Additionally, they use a variety of customer engagement and support software technologies such as Sentry, Ada, and Adobe Experience Platform.  For example, Ada is a technology used to provide chat bot capability for client support and can be identified by the presence of the string ``adaEmbed'' in the JavaScript files.
    \item The website relies on a handful of technologies related to Internet hosting such as Cloudflare, AWS, and the jQuery CDN, which can be identified by the fact that a file is accessed from the domain code.jquery.com during page load. This technique is therefore often able to uncover some fairly deep dependencies, e.g., the use of an Akamai-based bot identification technology is identified on this website via the presence of a single field in a cookie labelled ``ak\_bmsc''.
    \item The website also uses a variety of advertising platforms such as Amazon and Microsoft ads.
    \item Finally, the software stacks that this website is based on includes 22 different software components such as PHP, React, Drupal, and Ruby, as well as a host of JavaScript based technologies such as DataTables, Node.js, and Backbone.js.
    \item In total 55 distinct component technologies are identified.
\end{itemize}

As mentioned, the presence or absence of many of these technologies sheds light on an organization's cybersecurity posture, as it is generally indicative of technological choice, competence, and sophistication. Some are more directly connected to cybersecurity best practices. Thus, how an organization engages with these technologies holds power in estimating cyber risk. In the next section we detail how this raw data is converted into numerical features suitable for training a classifier.

\subsection{Feature extraction}\label{sec:extract}

We now describe the method we use to convert the raw data (technologies) captured by the crawls, which is returned in JSON format, into numerical features that can be fed into a supervised learning process to train a classifier.

First, we note that our raw JSON data includes not only the name of the technologies on the basis of their digital signatures, but also the specific version numbers of those technologies whenever they could be recognized. Additionally, each identified technology can be associated with one or more categories (provided by Wappalyzer) on the basis of the core functionality that the technology provides. This means that each category can represent potentially many possible versions/types. We use binary features to represent the presence or absence of each under each category, but also use a numerical feature to capture the total number of presences (number of ``1''s among all the binary features) for each category.

More specifically, we use a simple one-hot encoding technique to convert the presence and absence of these technologies into binary and numerical features as follows.

\begin{itemize}
    \item \textbf{Technology names/versions}: For each captured technology, we record its name and, if available, version information, specifically the major and minor version numbers (while excluding patch, build, and revision numbers). As an example, jQuery 1.13.2 is divided into jQuery, jQuery 1 and jQuery 1.13. We then use one-hot encoding to convert each name/version into a binary feature that marks its presence in the list of technologies captured from a website. These binary features are organized into categories as shown in \autoref{table:technologies}.
    \item \textbf{Category/meta-category}: We create a numerical (integer) feature for each category and meta-category by counting the number of technologies captured from a website that fall within that group.
\end{itemize}

We further prune the full feature set by eliminating those that are observed (i.e., non-zero) in less than 20 of our crawled websites, yielding in a total of 1,013 features. Some examples of pruned features include infrequent technology names (e.g., AccuWeather, ExoClick, and Zendesk Chat), and infrequent versions of specific technologies (e.g., AngularJS 1.3, WordPress 4.8, and Yoast SEO 3.0).\footnote{Note that for infrequent versions of frequent technologies (e.g., WordPress 4.8), the technology itself is still captured by our feature set through the respective technology name.}

The above process results in different number of features associated with each category and meta-category (the number in parenthesis) in \autoref{table:technologies}. For instance, the category ``Cookie compliance'' under the meta-category ``Security/Privacy'' has 20 features. This means that there are a total of 19 possible technology names/version captured in this category after pruning (19 binary features marking the presence or absence of each), plus 1 numerical feature which counts the number of ``1''s among the 19 binaries. Similarly, the meta-category ``Security/Privacy'' itself has 37 features, which includes the total number of binary features across all three of its categories (33), the 3 numerical features counting the number of ``1''s within each category, and the 1 numerical feature that counts the total number of ``1''s within the meta-category itself.

\begin{figure}[t]
    \includegraphics[width=0.75\linewidth]{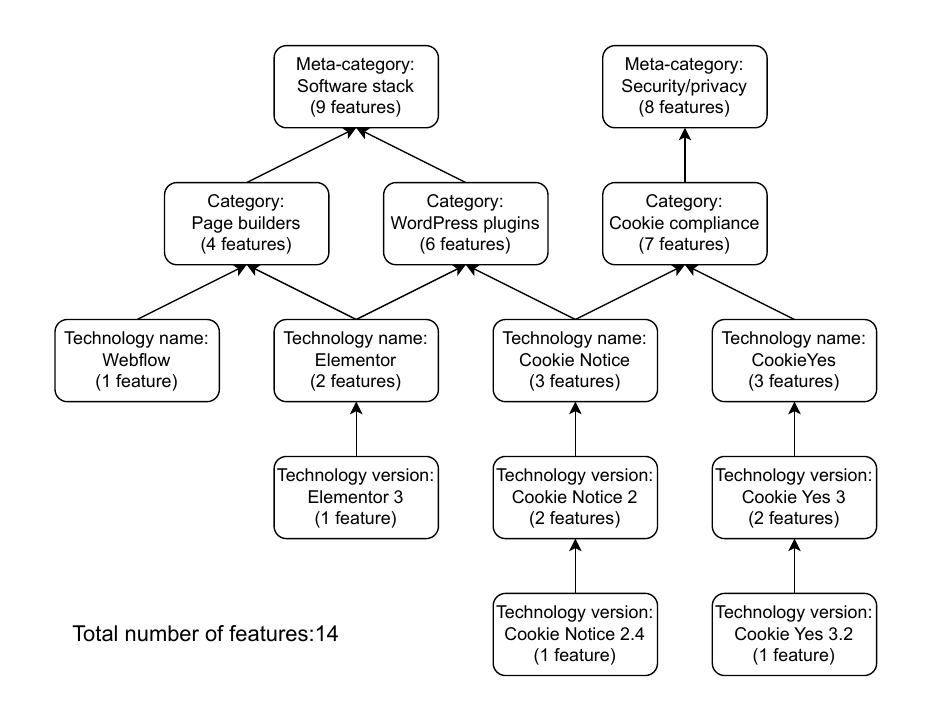}
    \caption{A partial hierarchy of technology names/versions, categories, and meta-categories. Each node represents a single binary (for technology names/versions) or numerical (for categories/meta-categories) feature. A technology can be attributed to multiple categories. We compute a node's associated number of features (in parentheses) by counting the number of child nodes with a directed path to the node, including the node itself. The same methodology is used to compute feature counts in \autoref{table:technologies} to avoid double counting.}
    \label{fig:tech-tree}
\end{figure}

Since a technology can potentially belong to multiple categories, there can be double counting of some binary features across different categories. These are removed when we tabulate the total number of features associated with a meta-category in \autoref{table:technologies}, resulting in a lower total than what might be suggested by those associated with each category. This is further illustrated in \autoref{fig:tech-tree}, where we show a partial example hierarchy with 9 technology names/version, 3 categories, 2 meta-categories, and a total of 14 features.

As mentioned earlier, a primary reason for utilizing Wappalyzer's categories and defining meta-categories of our own is to create groupings that bring together similar technologies/categories, and to create a distinction between categories from a risk management point of view. The above feature extraction process explicitly utilizes these groupings, by adding as a feature the counts of technologies present within each group. Note that a machine learning algorithm has no prior knowledge of similarities between technologies/categories. Therefore, these counts allow the learning algorithm to utilize such domain expertise. Additionally, they allow us to quantify a group's contribution to classifier outputs, as some can play a more important role in cyber risk assessment as we  demonstrate in \autoref{sec:performance}.

\subsection{Organization label data}\label{section:labels}

\begin{table}[t]
    \centering
    \caption{Summary of our dataset. Incident reports from VCDB/BFSR are first mapped to the victims' domains using a semi-automated Google search, followed by a web crawl to capture technology signatures. Non-incident (negative) samples are further refined by checking for the presence of a privacy policy to filter out non-organizational websites.}
    \label{table:dataset}
    \begin{tabular}{|c|c|c|cc|c|}
        \hline
        \multicolumn{2}{|c|}{\multirow{2}{*}{Source}} & \multicolumn{3}{c|}{Positives} & Negatives \\
        \cline{3-6}
        \multicolumn{2}{|c|}{} & VCDB & \multicolumn{2}{c|}{BFSR} & Tranco \\
        \hline
        \multicolumn{2}{|c|}{Year(s)} & 2022-23 & 2022-23 & 2024 & N/A \\
        \hline
        \multicolumn{2}{|c|}{\# of Incidents} & 859 & 997 & 416 & N/A \\
        \hline
        \# of & Initial & 830 & 959 & 407 & 10k \\
        domains & Valid crawl & 817 & 931 & 391 & 7,569 \\
         & Organizational & 817 & 931 & 391 & 3,971 \\
        \hline
        \multirow{2}{*}{Usage} & Training & Yes & Yes & No & Yes \\
        & Validation & Yes & Yes & Yes & Yes \\
        \hline
    \end{tabular}
\end{table}

The previous subsections describe how feature data is obtained for a given domain/URL. To train a supervised model to estimate cyber risk, we need to pair the feature data with a label, the occurrence of a cybersecurity incident. To do so we need example lists of organizations (with their corresponding Internet domains) with and without a reported incident. Below we detail how we acquire and sanitize this data to obtain high-quality ground-truth labels for model training. \autoref{table:dataset} summaries the curated datasets.

\subsubsection{Positive (incident) samples}\label{sec:positives}

We rely on two separate data sources to obtain a list of organizations that have experienced a data breach in the past. The first is the well-known VERIS Community Database (VCDB)~\cite{vcdb}.  This is a public database maintained by the Verizon RISK team capturing publicly disclosed security incidents, including those due to malware, hacking, social engineering (e.g., pretexting, phishing, scams, etc.), misuse (by an entrusted insider or partner), physical actions (e.g, theft, tampering, snooping, etc.), error, and environmental events (e.g, earthquakes, floods, power failures, etc.). We obtain all incident reports from the VCDB from 2022 and 2023 that are attributed to malware, hacking, and social engineering, which results in a total of 859 cybersecurity incidents.

As a second data source we focus on publicly reported ransomware incidents. There are several such listings available such as the state of ransomware reports maintained by BlackFog~\cite{blackfog}, an interactive map of ransomware attacks by StateScoop~\cite{statescoop}, the Critical Infrastructure Ransomware Attacks (CIRA)~\cite{rege2023free} dataset, and a crowdsourced ransomware payment tracker by Ransomwhere~\cite{cable2024ransomwhere}. For the purpose of this study we use BlackFog's 2022-2024 state of ransomware reports, referred to as BFSR for conciseness in the rest of our study. The processed BFSR data consists of 1416 ransomware incidents. From these, we use 997 incidents from 2022-2023 for training and evaluating our models, and provide an analysis on the remaining (held-out) 416 incidents between January and July 2024 later in \autoref{sec:result} (\autoref{fig:roc-blackfog}). From hereon, will explicitly use BFSR 22-23, BFSR 24, or BFSR 22-24 when referring to different subsets, or the entirety, of this dataset.

\paragraph{Domain mapping}: It is important to note that our positive samples only include the names of organizations with a reported cybersecurity incident. However, as described in \autoref{sec:method}, we need the corresponding Internet domains of these organizations to collect features for training and evaluating our classifier. In order to obtain the set of associated domains, we rely on a Google search based technique. Using a Google Custom Search JSON API\footnote{\url{https://developers.google.com/custom-search/v1/overview}} query, we first retrieve the primary Google search result for each cited organization name, excluding results that point to directories such as Wikipedia, LinkedIn, Bloomberg, Facebook/Instagram, etc. Next, we validate the result by inferring the organization name back from a web crawl of the retrieved domain and comparing it against the original name, as detailed below.

Using our web crawl agent, we parse the landing page of the associated website and extract the page title, meta description/keywords, OpenGraph meta tags (``og:title'', ``og:site\_name'', and ``og:description''), and the plain text HTML content (extracted using the Inscriptis library~\cite{weichselbraun2021}). We then provide these to an AI chatbot tasked with extracting the name of the underlying organization. We explicitly instruct the chatbot to prioritize the page title and copyright notice at the bottom of the page, falling back to analyzing the page content if those fail to identify the organization's name. We then use longest common substring matching to identify common substrings of three characters or more between the organization name cited in the incident report and the AI-extracted name, and compute a similarity score by dividing the total length of the matched substrings by the minimum length of the two names.

Finally, we manually inspect all samples with a similarity score below 0.9 (141 for VCDB and 390 for BFSR 22-24). Upon manual inspection, we found that our automated domain retrieval was 97.0\% accurate (with 26 corrections) for VCDB and 90.8\% accurate (with 130 corrections) for BFSR 22-24.\footnote{Those with similarly scores at or above 0.9 are $\sim$100\% accurate.} Note that for companies with multiple domains, our method selects the one with the highest Google search rank, which typically corresponds to their most frequently visited domain.

\paragraph{Positive labels dataset}: The incident description processing and domain mapping  technique described above  yields 830 unique domains from VCDB and 1,366 unique domains from BFSR 22-24. Note that the number of domains is lower than the original number of incidents because some domains are associated with multiple incident reports, as well as rare cases where we could not find a valid website for the cited organization name (9 for VCDB and 31 for BFSR 22-24).

The fact that we can find valid websites for $>$98\% of the incident reports highlights the potential of the proposed technique for producing data, features, and cyber risk assessment for almost all organizations regardless of their size, as opposed to prior techniques based on Internet scanning~\cite{liu2015cloudy} that fail to produce meaningful data and features for a large number of SMEs without an identifiable dedicated IP address space.

\subsubsection{Negative (non-incident) samples}\label{section:negatives}

In addition to the positive samples described above, we also obtain a set of negative samples by selecting 10k random domains from the Tranco top million list~\cite{pochat2018tranco}. In making this selection we employed a number of heuristics to filter out ``trivial'' negative samples. For instance, we intentionally exclude domains listed by Tranco that do not host a live website (e.g., non-web endpoints such as ``amazonaws.com''), and websites where no privacy policy could be identified -- the presence or absence of a privacy policy is used as a heuristic to help identify domains that belong to organizations rather than domains that are owned by individuals (e.g., blogs) or domains that do not represent a significant organization.

The purpose of these filtering processes is to ensure that we obtain ``high-quality'', comparable negative samples, i.e., they represent websites of entities/organizations that are subject to security incidents, but have not (to the best of our knowledge) experienced a data breach.

\paragraph{Negative labels dataset}: The processing and filtering technique described above results in a final dataset of 3,971 domains with a successful crawl and an existing privacy page. The next subsection describes our yield in more detail.

\subsection{Yield on feature and label data collection}\label{subsec:yield}

For a given domain, we try to construct a homepage URL using both HTTPS and HTTP protocols, with and without a "www" prefix (if the domain does not already include it). We then declare a success if any of these variations return an HTTP status code below 400, and at least one technology is captured by Wappalyzer. We obtain a successful crawl for 7,569 out of the 10k (75.7\%) of our non-incident (negative) domains described earlier, 817 out of 830 (98.4\%) of domains from VCDB, and 1,322 out of 1,366 (96.8\%) from BFSR 22-24. The lower number of successful crawls for the non-incident domains is largely due to Tranco including domains that do not host a website (e.g., non-web endpoints), while failures among VCDB and BFSR 22-24 are due to websites that block the crawler. We further filter non-incident domains by discarding those for which our crawler fails to navigate to a privacy page (with an HTTP status code below 400) as an additional attempt to eliminate non-organizational domains, yielding 3,971 domains (52.5\% of domains with a successful crawl) for our non-incident/negative population.  These numbers are summarized in \autoref{table:dataset}.

\section{Classifier Training and Output}\label{sec:result}
We next describe the training of a supervised model using the features and labels discussed in the previous section. We present how we train and evaluate the performance of our classifier model, examine its accuracy and robustness, and provide an analysis on the contribution of different feature groups to the classifier output (risk assessment on likelihood of cyber incidents).

\subsection{Model training}\label{sec:training}

We use gradient-boosted trees, specifically XGBoost~\cite{chen2016xgboost}, as our classification model due to the tabular nature of our data. Note that tabular data is defined as data that can be organized in rows and columns (similar to a spreadsheet), where each row represents an observation (i.e., a domain in our case), and each column represents a (categorical or numerical) feature. While deep learning models often achieve state-of-the-art performance on text and image data, tree-based models such as XGBoost frequently outperform neural networks on tabular data, especially for small and medium-sized ($<$10K samples) datasets~\cite{grinsztajn2022tree}. XGBoost is based on gradient boosting, where an ensemble of weak learners (decision trees) are trained in succession, with each tree correcting the errors made by its predecessors. While neural networks are prone to over-fitting on smaller datasets due to their high capacity, XGBoost incorporates various regularization techniques to reduce over-fitting, while also capturing complex feature interactions and providing interpretability through feature importances.

We use a $k$-fold cross validation, with $k = 5$, to partition our dataset into training/validation sets. More specifically, we divide our dataset into 5 equally sized subsets or \emph{folds}. We then train 5 classification models, each using 4 folds as the training set and the remaining fold as the validation/test set. By rotating the validation fold across these 5 iterations, we are able to generate a robust accuracy estimate by generating a classification output (also referred to as a \emph{score} throughout the discussion) for every sample in our dataset, whereas each sample is evaluated by a model that wasn't trained on it. An arbitrary sample can then be classified by averaging the outputs of all five trained models. $k$-fold cross validation is a commonly used approach that does not involve a fixed hold-out dataset for validation, and can produce more robust accuracy estimates for small datasets~\cite{hastie2009elements}. Nevertheless, we will use the more recent BFSR 24 dataset, which we completely exclude from model training, as a true held-out dataset for out-of-distribution evaluation in testing the robustness of our trained model later in this section.

We choose the following hyper-parameters to train our models, guided by cross-validation. We use a learning rate of 0.1, 1000 boosting rounds (resulting in an ensemble of 1000 trees), a loss guided grow policy for adding new nodes to a tree (i.e., splitting nodes with the highest change to the loss function), a maximum of 128 leaf nodes for each tree, and the histogram tree construction method with a maximum of 32 bins for bucketing continuous features. We train our models using the logistic regression objective, with the model outputting the probability that a given sample belongs to the positive class. We use early stopping to prevent over-fitting by halting the training process when the validation loss does not improve for 50 boosting rounds.

\subsection{Model performance}\label{sec:performance}

We now provide a detailed performance analysis of the trained models. We focus on three distinct performance criteria: model accuracy, feature importance, and model robustness.

\begin{figure}[t]
    \includegraphics[width=0.6\linewidth]{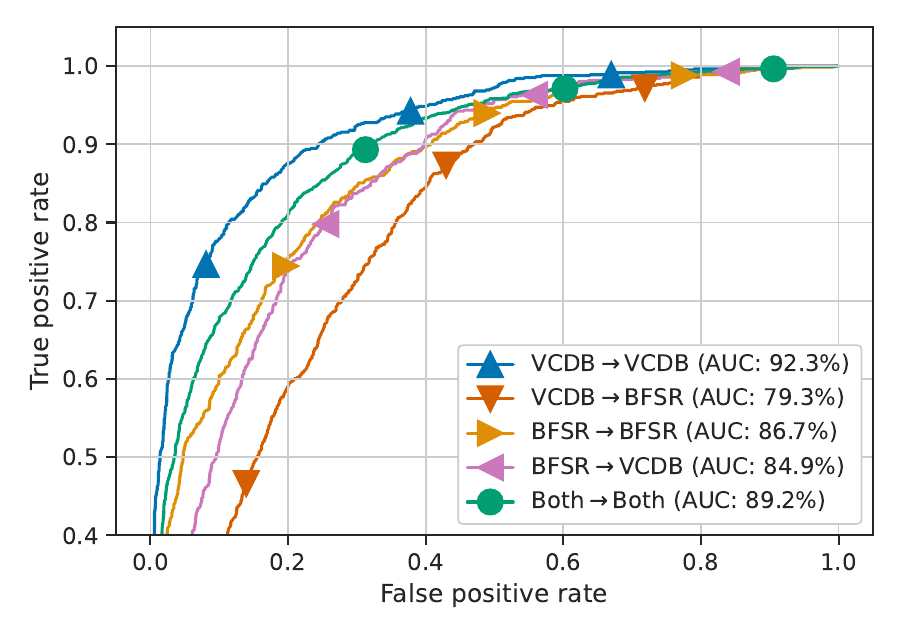}
    \caption{ROC curves of the trained classifier. $X \rightarrow Y$ refers to a model trained on $X$ and then evaluated on $Y$. The blue (orange) curve is trained and evaluated on VCDB (BFSR 22-23) labels only; the green  uses both sets of labels. The two worse-performing curves correspond to cross-dataset evaluation, the red trained on VCDB and evaluated on BFSR 22-23 and the purple trained on BFSR 22-23 and evaluated on VCDB.}
    \label{fig:roc}
\end{figure}

\subsubsection{Model accuracy}

The receiver operating characteristic (ROC) curve is a commonly used metric used to report the performance of a supervised learning model as it conveys the accuracy of a model's outputs across a range of potential operating conditions. \autoref{fig:roc} illustrates the ROC curve of model outputs and known true labels. Each point on the ROC curve represents a different threshold for converting continuous model outputs (i.e., the probability of a given sample belonging to the positive class) into distinct binary labels, demonstrating the trade-off between the true positive (TP) rate and the false positive (FP) rate across various classification thresholds. The area under the ROC curve (ROC AUC or simply AUC) summarizes a model's performance in a single metric, and can be interpreted as the probability that a randomly chosen positive sample will receive a higher score (risk estimate) than a randomly chosen negative sample.

\autoref{fig:roc} shows five distinct variations of the model. Three of them are models trained using different sets of positive labels from (1) VCDB, (2) BFSR 22-23, and (3) the combination of both, respectively. The remaining two demonstrate cross-dataset results. The first three models, using consistent training-testing labels (the blue, orange, and green curves), achieve AUCs of 92.3\% for VCDB, 86.7\% for BFSR 22-23, and 89.2\% for the combined dataset. For all three models, a good operating condition (the ``knee'' of the ROC) has a FP rate of 20\% and a TP rate between 75.5\% (for BFSR 22-23) and 87.5\% (for VCDB). It is also possible to choose a lower FP of 10\%, yielding a TP between 60.5\% (for BFSR 22-23) and 78.0\% (for VCDB).

\autoref{fig:roc} also contains two cross-dataset results: a model trained on VCDB but evaluated using BFSR 22-23 (red) and vice versa (purple). Although each model remains meaningful (i.e., still outperforming random guessing, which would yield a 50\% AUC) when applied to the alternate dataset, their performance is clearly inferior to the first three models trained and evaluated with similar types of labels. Specifically, the model trained on VCDB achieves an AUC of 79.3\% on BFSR 22-23, while the one trained on BFSR 22-23 obtains an AUC of 84.9\% on VCDB, both lower than their respective within-dataset AUCs. This drop in performance indicates notable differences in the underlying distributions of VCDB and BFSR 22-23. It also underscores the importance of using more comprehensive training data if the goal is to generalize to a broad spectrum of cyber incidents. This is further inspected and discussed in \autoref{sec:discussion}.

\begin{figure}[t]
    \includegraphics[width=0.6\linewidth]{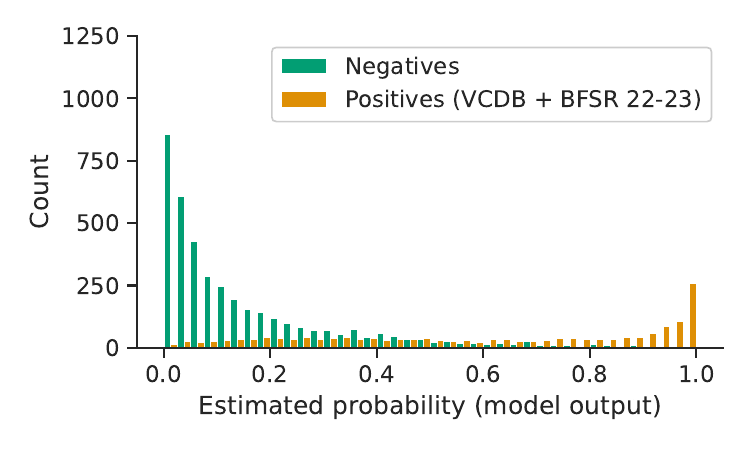}
    \caption{Distribution of classifier outputs on all positive and negative samples used in this study, including both VCDB and BFSR 22-23 samples. We observe a clear distinction where positive (negative) samples are concentrated on the right (left).}
    \label{fig:dist}
\end{figure}

As the models are trained with a positive label value of 1 indicating a cybersecurity incident and a negative label value of 0 indicating a lack of a cybersecurity incident, the model output will be higher (on a scale of 0 to 1) for samples that more closely match the features associated with a positive label than one whose features more closely match with negative labels. \autoref{fig:dist} shows an alternative way to visualize the accuracy of a model by illustrating its ability to separate positive and negative samples. The figure shows the model output distribution for the combined model for all known positive and negative labels. We can clearly see that for the positive label population, the model outputs are clustered at the higher end of the output range, while for known negative labels they are  clustered towards the lower end of the range.

\begin{figure*}[ht]
    \begin{tikzpicture}
        \node(contribs)[anchor=south west,inner sep=0pt]{\includegraphics[width=\linewidth]{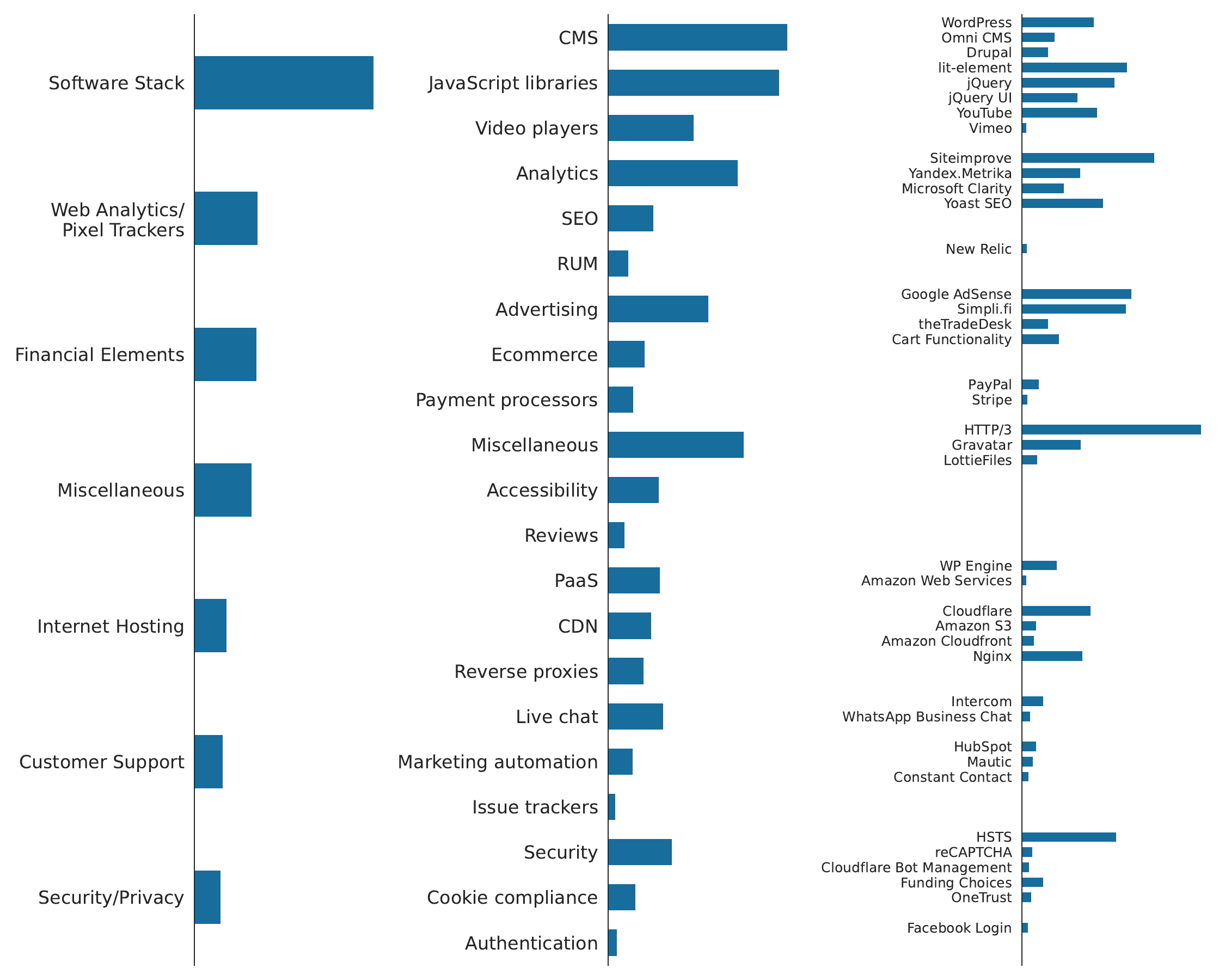}};
        \begin{scope}[x={(contribs.south east)},y={(contribs.north west)}]
            \foreach \i/\j in {{(0.01, 0.1540)/(0.99, 0.1540)},{(0.01, 0.2924)/(0.99, 0.2924)},{(0.01, 0.4308)/(0.99, 0.4308)},{(0.01, 0.5693)/(0.99, 0.5693)},{(0.01, 0.7077)/(0.99, 0.7077)},{(0.01, 0.8461)/(0.99, 0.8461)}}
                \draw[black] \i -- \j;
            \foreach \i/\j in {{(0.33,0.0616)/(0.99,0.0616)},{(0.33,0.1077)/(0.99,0.1077)},{(0.33,0.2000)/(0.99,0.2000)},{(0.33,0.2462)/(0.99,0.2462)},{(0.33,0.3385)/(0.99,0.3385)},{(0.33,0.3846)/(0.99,0.3846)},{(0.33,0.4769)/(0.99,0.4769)},{(0.33,0.5231)/(0.99,0.5231)},{(0.33,0.6154)/(0.99,0.6154)},{(0.33,0.6615)/(0.99,0.6615)},{(0.33,0.7538)/(0.99,0.7538)},{(0.33,0.8000)/(0.99,0.8000)},{(0.33,0.8923)/(0.99,0.8923)},{(0.33,0.9384)/(0.99,0.9384)}}
                \draw[black, densely dotted] \i -- \j;
        \end{scope}
    \end{tikzpicture}
    \caption{Contributions of different feature groups to model outputs: meta-categories (left), categories (middle), and technologies (right). We report the top 3 contributing categories within each meta-category, and the top 3 contributing technologies within each category. The Communication Systems meta-category has a negligible contribution and is therefore not shown.}
    \label{fig:contribution}
\end{figure*}
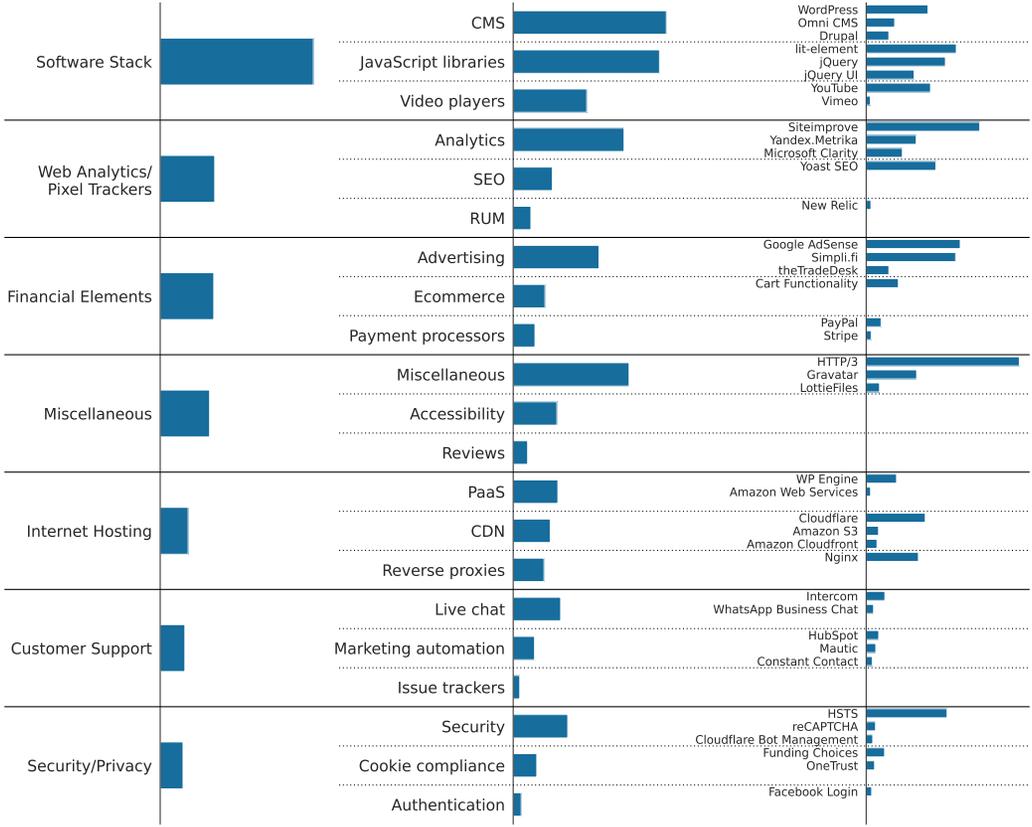

\subsubsection{Feature importance}

As shown in \autoref{table:technologies} we derive and use a large number of binary and integer features for the purpose of building our models. To understand the importance of different features and feature groups for estimating risk, we use SHAP values~\cite{lundberg2017unified} to quantify feature contributions. For any given sample (i.e., domain) a SHAP value is assigned to each of the 1,013 features, where a positive (negative) SHAP value denotes that the associated feature is driving the model's output toward the positive (negative) class, with higher absolute values indicating a stronger effect. We use SHAP values due to their desirable properties such as local accuracy (attributions sum up to the output of the model), missingness (missing features are given no importance), and consistency (modifying a model so that a feature is given more weight never decreases its attribution). We quantify the contribution of a given category or meta-category to the model's output as the sum of contributions from all individual features under that group. Since SHAP value are assigned to individual samples, we quantify the overall contribution of a feature (or feature group) to the model as as the mean absolute value of all corresponding SHAP value.

\autoref{fig:contribution} summarizes some of the most significant contributing features for our combined model. As we do not bias our features in anyway in the training, the relative importance is an outcome of the training process and is a direct reflection of the input data. The figure helps to provide an intuitive understanding of why we might expect the model output to be high or low depending on our understanding of the input features. \autoref{fig:contribution} shows the most relevant features in the form of the technology hierarchy: from the finest level of granularity to the ultimate meta-categories at the coarsest level. We highlight some of the more noteworthy observations below.

\begin{enumerate}
    \item The most influential meta-category for the model is the Software Stack, followed by the Web Analytics/Pixel trackers category.
    \item Further within the Software Stack meta-category, items categorized as CMS (content management systems) and JavaScript libraries are the most relevant.
    \item Finally, at the finest level of granularity, we see that CMS technologies such as WordPress, Omni CMS, and Drupal, and JavaScript libraries such as jQuery and lit-element are the most influential in determining model outputs.
    \item Similarly, we find that the use of web analytics technologies such as Siteimprove and Yoast SEO, advertising and financial systems such as Google AdSense and Simpli.fi, web hosting technologies such as Cloudflare and Nginx, and security-related technologies such as the use of HSTS are the most influential in determining model outputs.
\end{enumerate}

A feature (or feature group) can influence the model output in either a positive (pushing the output toward a positive label) or negative manner.  For instance, on average, HTTP/3 and Cloudflare are technologies that are relatively significant positive contributors to lower estimated risk, meaning that their presence (absence) leads to lower (higher) estimated risk, whereas Siteimprove and Simpli.fi are relatively significant negative contributors, meaning that their presence (absence) leads to higher (lower) estimated risk.

\begin{figure}[t]
    \includegraphics[width=0.6\linewidth]{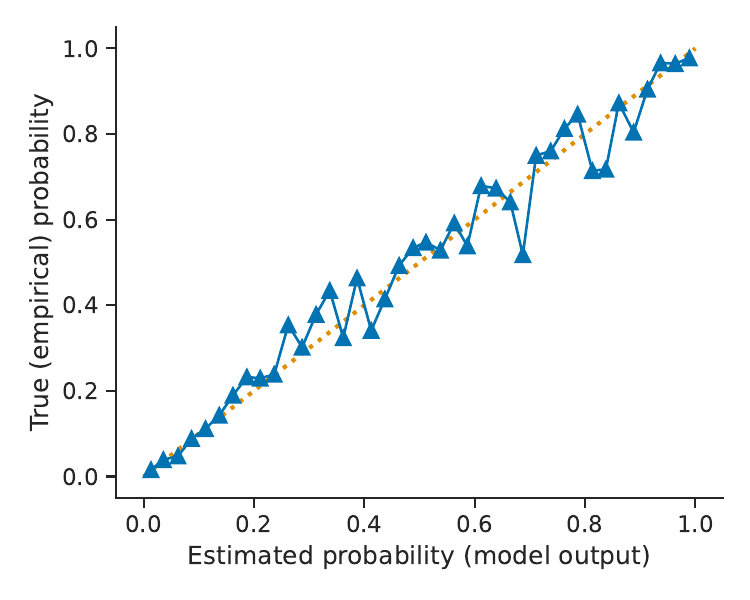}
    \caption{Binned model outputs (risk estimates) vs. true empirical probabilities of belonging to the positive (incident) class. The dotted straight line represents a perfectly calibrated model, showing that our model outputs are well-calibrated.}
    \label{fig:output}
\end{figure}

\subsubsection{Model robustness}

One of the most important criteria for an effective model is its robustness to changes in input datasets. As already discussed, \autoref{fig:roc} shows robustness to variations in input training data.  In particular, the combined model demonstrates this resilience. A second aspect of robustness is the extent to which model calibration is needed post-training in order for the  output probabilities to match the empirical probabilities calculated directly from the input datasets.

\autoref{fig:output} shows how raw classifier outputs from the combined model line up with respect to the true (empirical) probability of a sample belonging to the positive (incident) class. These plots are generated by first binning model outputs into 40 uniform bins, computing the empirical probability by taking the ratio of positive labels to the total number of samples in each bin (y-axis), and then plotting that against the estimated probability (x-axis)~\cite{niculescu2005predicting}. The resulting calibration plot is then compared against the 45-degree line $y = x$, representing a perfect model where the estimated probability matches exactly the true empirical probability. When the two are not well matched, practitioners typically add a model calibration step, where a function is introduced to map the model output to the line $y = x$. As can be seen, the output of our classifier is extremely well behaved as to effectively requiring no further calibration.

\begin{figure}[t]
    \includegraphics[width=0.6\linewidth]{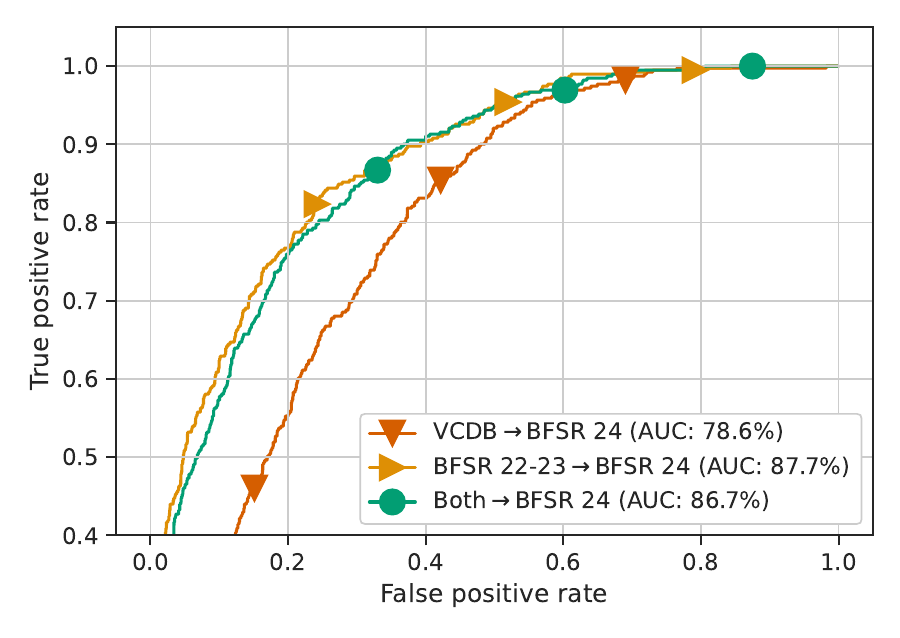}
    \caption{Accuracy of models for scoring BFSR 24 incidents. Models trained on BFSR 22-23 and the combination of VCDB and BFSR 22-23 maintain a high performance. A model trained only on VCDB performs poorly (similar to \autoref{fig:roc}) while still outperforming random chance.}
    \label{fig:roc-blackfog}
\end{figure}

The last aspect of model robustness we examine is how model performance degrades with time lapse between the last training sample and new testing data.  While we do not have any label data in 2024 for the VCDB dataset, we do have 391 positive samples (with valid crawls) for the BFSR 24 dataset starting January through July 2024. \autoref{fig:roc-blackfog} shows the performance of our trained models on this more recent dataset. The AUC shows the same level of performance on this newer dataset, indicating very robust performance over a significant period of time. We see a similar performance degradation when using a model trained on one label set but tested on the other, as we have seen earlier in \autoref{fig:roc}. Thus, our models are highly robust in time, but it is important to train a model on the intended incident types. More on this is discussed in \autoref{sec:discussion}. 

\section{Enhancements with Additional Features}\label{sec:enhancement}
The model presented in the previous section is trained entirely on publicly available, crawled website data. We have shown the significant power embedded in this type of data in estimating cyber risks. At the same time. there is of course an array of auxiliary information that could potentially be tapped into as additional features in training an ``augmented'' model. One possibility is information one could gather from domain related DNS data; another possibility is the option of deriving structural features from websites such as number of links, images and layout; yet another is the possibility of using industry sector information associated with a given domain/organization. We will take a closer look at this last option in this section. We first describe how we obtain this information and then discuss its impact on model performance.

\begin{figure}[t]
    \includegraphics[width=0.75\linewidth]{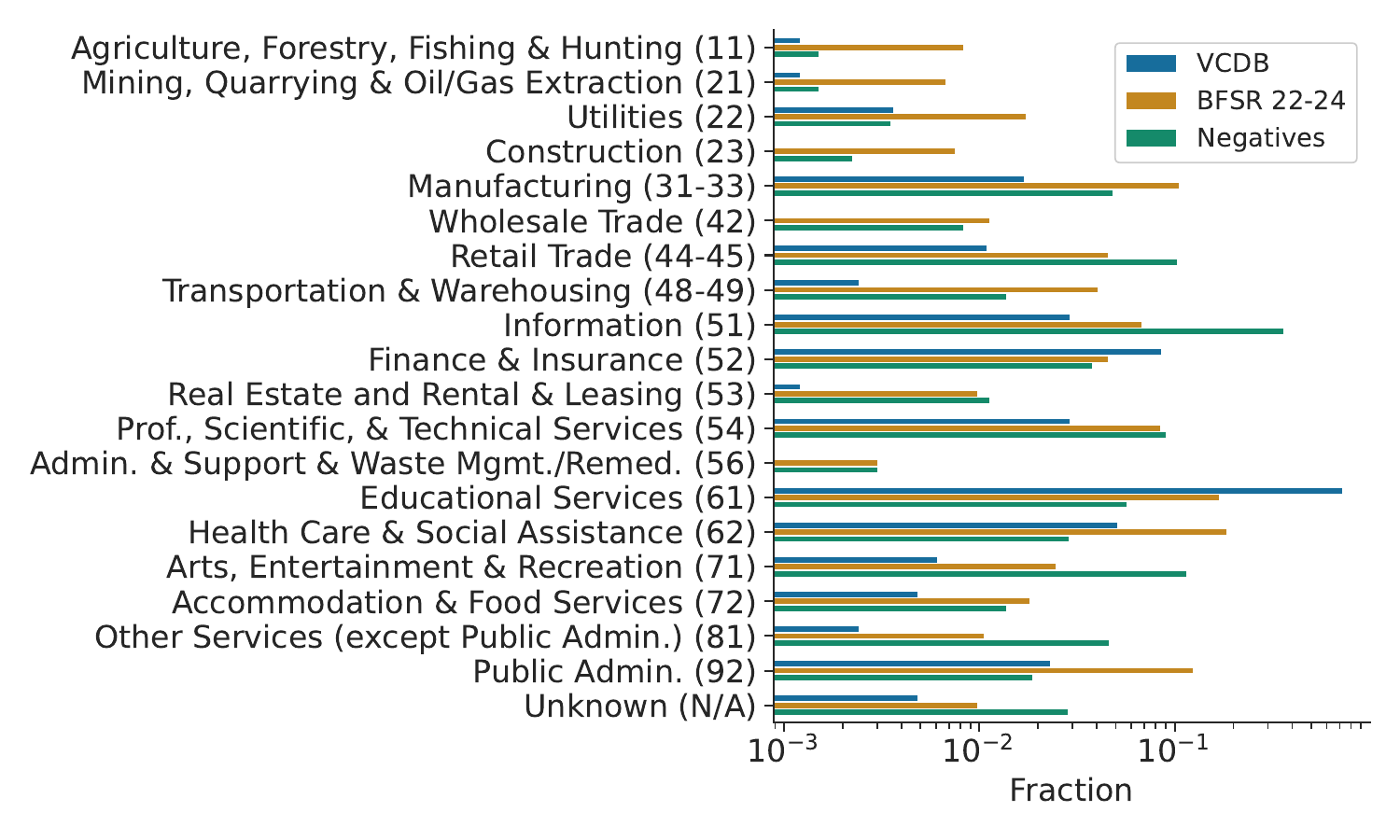}
    \caption{Distribution of AI-derived sectors for positives (VCDB and BFSR 22-24) and negatives. Numbers in parentheses correspond to the two-digit NAICS codes. A log scale is used for the x-axis.}
    \label{fig:sectors}
\end{figure}

The business sector of an organization can have a significant impact on its cyber risk, as different industries can face varying levels and types of threats. As an example, according to the IBM X-Force Threat Intelligence Index 2024~\cite{xforce}, manufacturing was the most attacked industry in 2023, followed by finance and insurance in second place, and professional, business, and consumer services in third place. In all three industries, installing malware and ransomware were the most common actions by cyber criminals. This means that sector information can hold significant predictive power in estimating risks. One way of obtaining this information is by purchasing specialty data or subscribing to services such as Dun \& Bradstreet.\footnote{\url{https://www.dnb.com}} However, we use an alternate approach, similar to the one described in \autoref{sec:positives}, by using an AI chatbot to determine a website's associated sector, with the knowledge that the homepage of an organization contains crucial information that can be used to identify its business sector. We feed the same page metadata and plain text content to the chatbot, instructing it to identify and map the organization's business sector to one of the 20 2-digit sectors defined by the North American Industry Classification System (NAICS)~\cite{naics}. While this mapping is not perfect,\footnote{As an example, the chatbot categorizes a number of health insurance companies as ``Health Care \& Social Assistance'', while ``Finance \& Insurance'' is often reported for these companies.} we observe that the resulting data is still useful in that it substantially improves the performance of our models as we demonstrate shortly. \autoref{fig:sectors} displays the distribution of the resulting sectors for each of our datasets, where we see a clear distinction between their demographics. In particular, VCDB incidents are heavily biased toward educational services, while the information sector is more frequently observed in our negatives (non-incident) set. BFSR 22-24 exhibits more diversity with health care, education service, and public administration being the most observed sectors.

\begin{figure}[t]
    \includegraphics[width=0.6\linewidth]{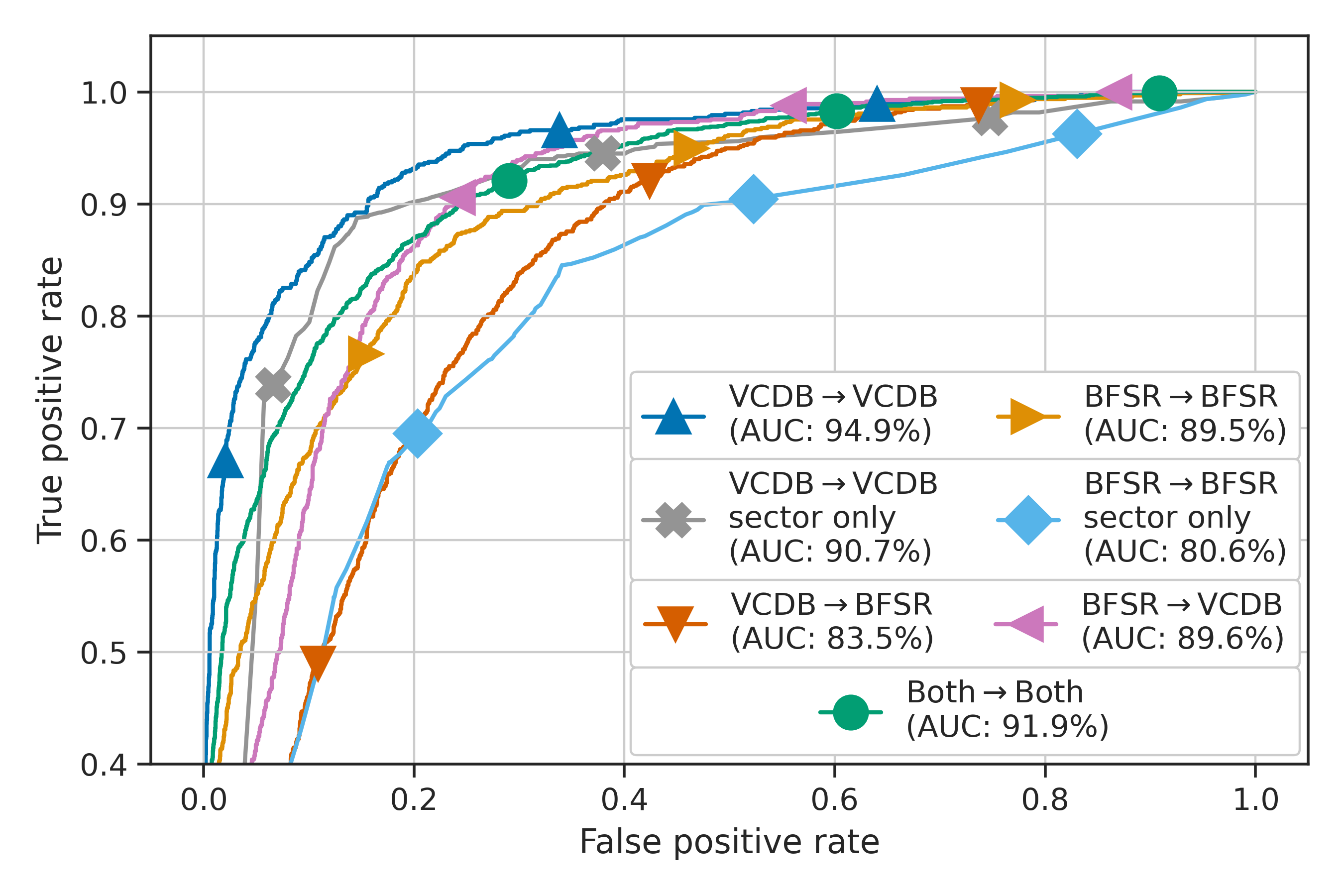}
    \caption{Accuracy of models trained on the combination of technology features and sector information (as an additional feature). We use the same training/evaluation scheme as in \autoref{fig:roc} with two additional curves using only sector information as features (i.e., excluding technology features): one trained and evaluated on VCDB (gray), the other trained and evaluated on BFSR 22-23 (light blue).}
    \label{fig:roc-sector}
\end{figure}

The performance of trained models with sector information as an extra one-hot encoded feature is given in Figure \ref{fig:roc-sector}, showing a clear performance improvement across the board over \autoref{fig:roc}. Interestingly, we observe that the addition of sector also improves cross-dataset performance.

The figure also includes two additional curves trained only on sector information. We observe a much higher performance for VCDB (90.7\% AUC) than BFSR 22-23 (80.6\% AUC), which is attributed to the high presence of educational services in VCDB. While using sector information alone appears to produce  decent performance (for VCDB), the presence of technology features significantly improves performance for both datasets when compared to models trained solely on sector data. Notably, the AUC increases (from 90.7\% to 94.9\% for VCDB, and from 80.6\% to 89.5\% for BFSR 22–23) represent substantial improvements by reducing the error by more than 40\%. This underscores the importance of incorporating technology features to produce accurate and robust risk estimates.

\section{Discussion}\label{sec:discussion}
\subsection{The difference between incidents broadly defined and the more specific ransomware attacks}

\begin{figure}[t]
    \begin{subfigure}[t]{0.35\linewidth}
        \includegraphics[width=\linewidth]{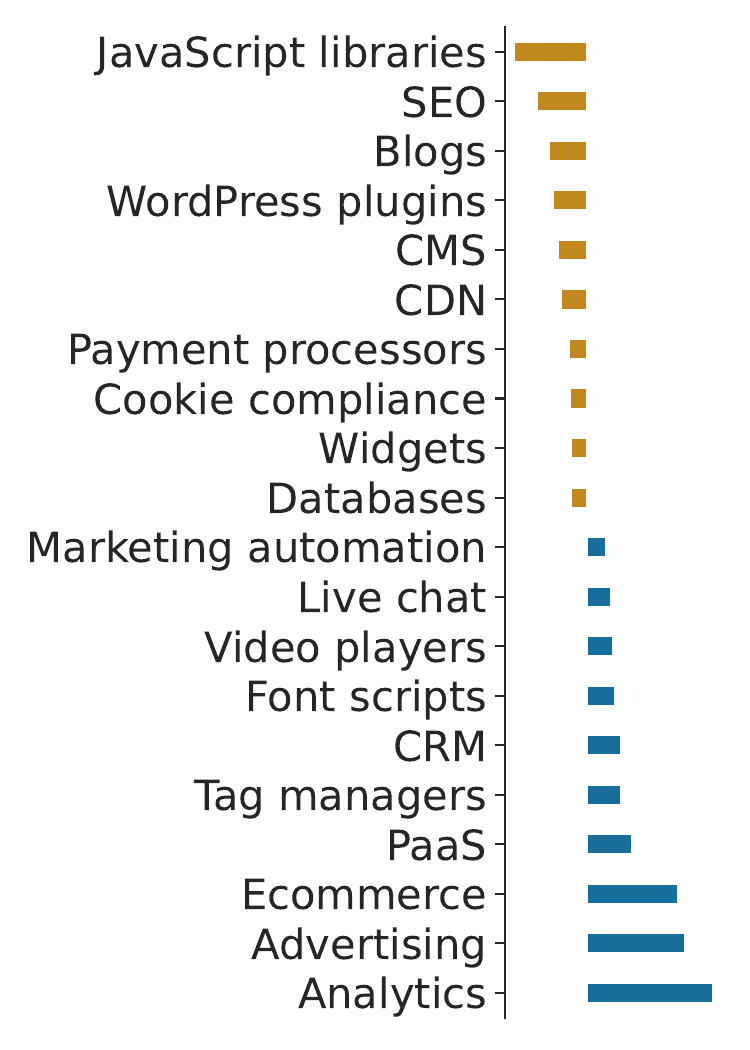}
        \caption{Categories}
    \end{subfigure}
    \begin{subfigure}[t]{0.35\linewidth}
        \includegraphics[width=\linewidth]{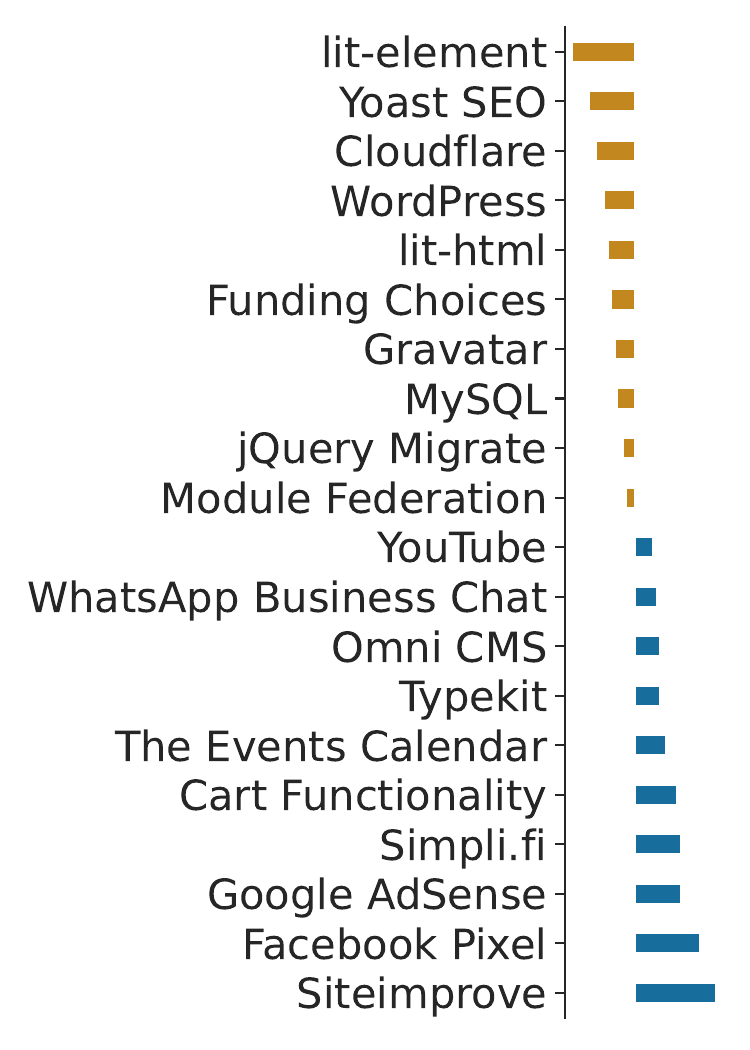}
        \caption{Technologies}
    \end{subfigure}
    \caption{Technology categories (left) and names (right) with the largest differences in contributions for the VCDB and BFSR 22-23 models. Positive (blue) bars indicate a higher contribution to the VCDB model, while negative (orange) bars indicate a higher contribution to the BFSR 22-23 model.}
    \label{fig:feature-comparison}
\end{figure}

We see from both Figures \ref{fig:roc} and \ref{fig:roc-blackfog} that training on one label dataset (e.g., VCDB) but evaluating on another results in significant performance degradation (e.g., AUC of 79.3\% for BFSR 22-23 in \autoref{fig:roc} and 78.6\% for BFSR 24 in \autoref{fig:roc-blackfog}). These numbers suggest that the fact that the validation is done using a separate, out-of-distribution hold-out set (BFSR 24) is only a minor contributor to the performance degradation, and that the main reason lies in inherent differences between the general population of cyber crime victims (VCDB) and the more specific ransomware victims (BFSR). The respective demographics of the two populations is already shown in \autoref{fig:sectors}; here we further examine the difference in feature importance between models trained on VCDB and BFSR 22-23, respectively (\autoref{fig:feature-comparison}).

More specifically, \autoref{fig:feature-comparison} displays the the top 20 categories and technologies with the largest difference in model contribution, where positive (blue) values indicate a higher contribution to the VCDB model; negative (orange) values indicate a higher contribution to the BFSR 22-23 model. We see the VCDB model's emphasis on technologies in the Analytics, Advertising, and Ecommerce categories (Financial Elements meta-category for the latter two), while the BFSR 22-23 model places greater importance on technologies in the JavaScript libraries and SEO categories. These results highlight the importance of ensuring a diverse set of training data when constructing generalized models, and the importance of choosing a representative label dataset for such cases, as differences in victim demographics can degrade performance when training and evaluation data are misaligned. As results in \autoref{fig:roc-sector} clearly indicate, it is possible to build high-performance special purpose models for targeted use cases.

\subsection{Comparison with study in \cite{liu2015cloudy}}

The overall goal of our study and the one presented in \cite{liu2015cloudy} are the same; however, we rely on entirely different data to arrive at similar results.  While we are unable to compare the results of our study with those presented in \cite{liu2015cloudy} on a organization-by-organization basis, we note that they are based partially on the same label dataset (VCDB). The accuracy of the models indicates a high degree of statistical alignment in the outputs of the two models, even though they use very different data (technologies used by an organization vs. host/network management and botnet activities).  It is also important to note that in each study the relationship between the features and the classifier output is correlational rather than causal.

As already mentioned, the biggest advantage of the present approach (and our main motivation) is getting around the challenges in using host-level (scan) data and the IP address attribution that it necessitates.  There is an additional difference in data types:  some of the data used in \cite{liu2015cloudy} (i.e., the reputation blacklists, or RBLs) is in the form of time series and \cite{liu2015cloudy} was able to align its feature and label data in time by collecting features prior to the known incident date to perform forecasting; our data is not aligned in the same way and for this reason we have avoided using the term ``forecast''. However, our trained classifier is indeed a predictor in the classical machine learning sense. In the remainder of this section, we provide additional experiments using historical archived crawl data to explore a predictive setting.

Similar to \cite{liu2015cloudy}, our positive and negative samples are drawn from different pools: the former from a breach dataset biased toward larger organizations (and those associated with sectors/regions with higher reporting requirements), and the latter from a broader pool. As noted by \cite{woods2021sok}, this can potentially lead the model to distinguish primarily between large corporations and smaller (or non) organizations. We have tried to mitigate this by excluding websites lacking privacy policies as mentioned earlier, thereby filtering out non-organizational web servers. At the same time, it is worth noting that cyber risk itself is inherently biased: companies of a certain size and/or in a certain industry sector are more likely to be targeted than others; size and sector also contribute to the nature of their digital assets and how difficult it is to protect them. In this sense the model is not trained to exclude this bias; however, the fact that there is substantial risk separation among firms of the same type is indicative of the model successfully capturing more risk signals than just size or sector. From a practical standpoint, an entity may be more interested in assessing its \emph{relative} risk by comparing with its peer group (same type/sector, etc.), which can be easily accomplished using our model output.

\subsection{Using historical data for training} As previously mentioned, our crawled data is not temporally aligned with incident dates, since the crawls occur after these incidents have taken place. To simulate a predictive scenario, we leverage the Wayback Machine~\cite{wayback} to retrieve historical snapshots of our crawled pages as follows. For positive (incident) samples, we select the most recent snapshot of a URL captured before the reported incident date. For negative samples, we (1) define a start date as the later of either the earliest available snapshot for the corresponding website or January 1, 2022, (2) uniformly sample a date between this start date and December 31, 2023, and (3) retrieve the most recent snapshot taken prior to the sampled date for each URL that our crawler visited. We subsequently employ a Go-based implementation of Wappalyzer\footnote{\url{https://github.com/projectdiscovery/wappalyzergo}} to perform technology detection on the archived pages. Using this process, we capture at least one technology using historical snapshots for 3,531/3,971 (88.9\%) of our non-incident (negative) samples, 805/817 (98.5\%) of VCDB samples, and 895/931 (96.1\%) of BFSR 22-23 samples. Across all URLs visited by our crawler, we obtain historical snapshots for 75\% and 67\% of the URLs corresponding to positive and negative samples, respectively. While the Wayback Machine frequently archives a website's homepage, it is more likely to miss subpages (particularly for less popular websites) leading to reduced coverage. Across all samples, the median time between the archived capture and the queried date (i.e., the incident date for positive samples or the sampled date for negative samples) is 33 days, with 20\% of snapshots taken more than 90 days away from the queried date.

\begin{figure}[t]
    \includegraphics[width=0.6\linewidth]{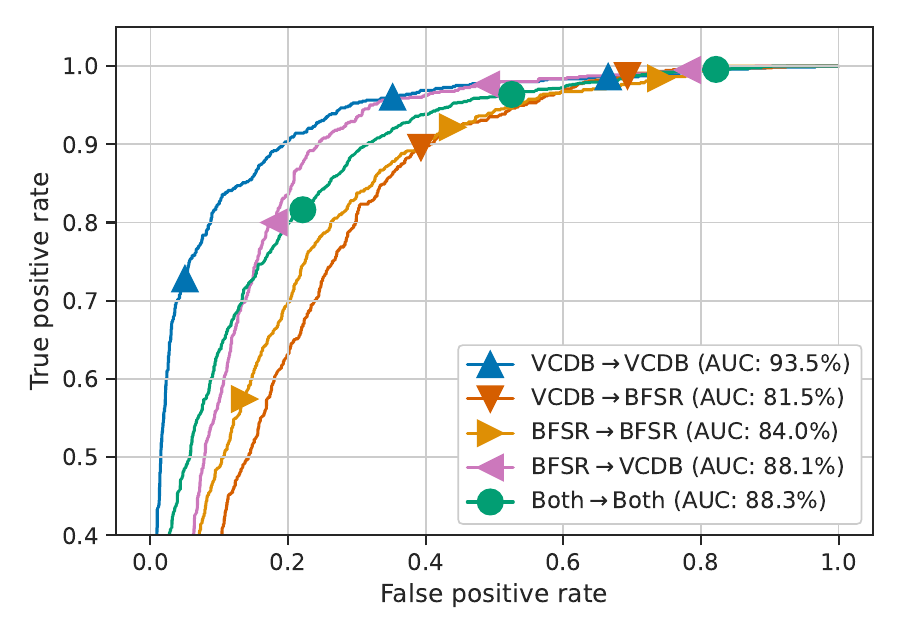}
    \caption{Accuracy of models trained on combined technology features from Wayback Machine data and sector information. Compared with \autoref{fig:roc-sector}, we observe a drop in performance across the board, which we attribute to incomplete coverage of website technology stacks in Wayback Machine data.}
    \label{fig:roc-sector-wayback}
\end{figure}

Next, we train and evaluate classifiers using the same experimental setup as in \autoref{fig:roc-sector}, using the combination of technology features obtained from Wayback Machine data and sector information. Our results are presented in \autoref{fig:roc-sector-wayback}, where the sector-only curves from \autoref{fig:roc-sector} are omitted to avoid repetition. Comparing the two figures, we observe an overall decline in performance, with the largest drop in performance occurring in the prediction of BFSR 22-23 incidents. We attribute this to limitations of Wayback Machine data. Specifically, many technologies on a webpage are detected by examining additional resources loaded in a browser, resources not necessarily captured in Wayback Machine snapshots which provide only the static HTML content. Combined with lower coverage on subpages (beyond the homepage), these factors lead to suboptimal technology detection from historical data. Consequently, while our feature exaction yields a total of 1,013 technology features from live crawls, this is reduced to 590 from Wayback Machine snapshots. Training classifiers on this reduced feature set even for live crawl data results in nearly the same level of performance degradation as seen in \autoref{fig:roc-sector-wayback}. However, while the total number of detected technologies using historical data is substantially lower, on average 80.3\% (and a median of 87.5\%) of technology names detected on a website using historical data also appear in our live crawls, suggesting that the set of technologies deployed on a website remains fairly stable over time. Therefore, we note that a more precise and scientific way of performing prediction is through the accumulation of live crawls, not by using Wayback Machine data due to the latter's limitations pointed out above. We are not yet able to perform this exercise due to limited historical depth in our dataset at this point, but continued data collection is an important part of our effort going forward that will eventually allow us to train our models in a truly predictive setting.

\section{Conclusions and Future Work}\label{sec:conclusion}

This paper presented a novel approach to building cybersecurity risk assessment models for organizations, using information and digital signals derived directly from the website associated with these organizations.  This approach has significant advantages over previous approaches that rely heavily on Internet scans and other data that can only be associated with IP addresses. The mappings from IP addresses to organizations can be extremely error prone and is entirely non-existent for millions of organizations that do not have dedicated and identifiable IP address-based assets. The model presented in this paper circumvents these limitations and can evaluate an order of magnitude larger number of organizations, as it only requires web presence for acquiring features. Our future work includes continued  collection of both feature and label data to enable forecasting of future incidents. We will also explore other auxiliary information such as DNS records and firmographics (e.g., employee count) to further enhance the model performance.

\bibliographystyle{plain}
\bibliography{references}

\begin{thebibliography}{10}

\bibitem{antonakakis2017understanding}
Manos Antonakakis, Tim April, Michael Bailey, Matt Bernhard, Elie Bursztein, Jaime Cochran, Zakir Durumeric, J~Alex Halderman, Luca Invernizzi, Michalis Kallitsis, Deepak Kumar, Chaz Lever, Zane Ma, Joshua Mason, Damian Menscher, Chad Seaman, Nick Sullivan, Kurt Thomas, and Yi~Zhou.
\newblock Understanding the {Mirai} botnet.
\newblock In {\em USENIX Security Symposium}, pages 1092--1110, 2017.

\bibitem{bano2018scanning}
Shehar Bano, Philipp Richter, Mobin Javed, Srikanth Sundaresan, Zakir Durumeric, Steven~J Murdoch, Richard Mortier, and Vern Paxson.
\newblock Scanning the {Internet} for liveness.
\newblock {\em ACM SIGCOMM Computer Communication Review}, 48(2):2--9, 2018.

\bibitem{bilge2017riskteller}
Leyla Bilge, Yufei Han, and Matteo Dell'Amico.
\newblock {RiskTeller}: Predicting the risk of cyber incidents.
\newblock In {\em ACM SIGSAC Conference on Computer and Communications Security}, pages 1299--1311, 2017.

\bibitem{blackfog}
{Blackfog}.
\newblock The state of ransomware 2024.
\newblock \url{https://web.archive.org/web/20250219091716/https://www.blackfog.com/the-state-of-ransomware-2024}.

\bibitem{cable2024ransomwhere}
Jack Cable.
\newblock Ransomwhere: A crowdsourced ransomware payment dataset.
\newblock \url{https://doi.org/10.5281/zenodo.13999026}, 2024.

\bibitem{chen2016xgboost}
Tianqi Chen and Carlos Guestrin.
\newblock {XGBoost}: A scalable tree boosting system.
\newblock In {\em ACM SIGKDD International Conference on Knowledge Discovery and Data Mining}, pages 785--794, 2016.

\bibitem{phishtank}
{Cisco Talos Intelligence Group}.
\newblock Phishtank.
\newblock \url{https://web.archive.org/web/20250519002908/https://www.phishtank.com}.

\bibitem{demarinis2019scanning}
Nicholas DeMarinis, Stefanie Tellex, Vasileios~P Kemerlis, George Konidaris, and Rodrigo Fonseca.
\newblock Scanning the {Internet} for {ROS}: A view of security in robotics research.
\newblock In {\em International Conference on Robotics and Automation}, pages 8514--8521. IEEE, 2019.

\bibitem{durumeric2015search}
Zakir Durumeric, David Adrian, Ariana Mirian, Michael Bailey, and J~Alex Halderman.
\newblock A search engine backed by {Internet}-wide scanning.
\newblock In {\em ACM SIGSAC Conference on Computer and Communications Security}, pages 542--553, 2015.

\bibitem{durumeric2014matter}
Zakir Durumeric, Frank Li, James Kasten, Johanna Amann, Jethro Beekman, Mathias Payer, Nicolas Weaver, David Adrian, Vern Paxson, Michael Bailey, and J.~Alex Halderman.
\newblock The matter of {Heartbleed}.
\newblock In {\em Internet Measurement Conference}, pages 475--488. ACM, 2014.

\bibitem{durumeric2013zmap}
Zakir Durumeric, Eric Wustrow, and J~Alex Halderman.
\newblock {ZMap}: Fast {Internet}-wide scanning and its security applications.
\newblock In {\em USENIX Security Symposium}, pages 605--620, 2013.

\bibitem{felt2017measuring}
Adrienne~Porter Felt, Richard Barnes, April King, Chris Palmer, Chris Bentzel, and Parisa Tabriz.
\newblock Measuring {HTTPS} adoption on the web.
\newblock In {\em USENIX Security Symposium}, pages 1323--1338, 2017.

\bibitem{feng2018acquisitional}
Xuan Feng, Qiang Li, Haining Wang, and Limin Sun.
\newblock Acquisitional rule-based engine for discovering {Internet-of-Things} devices.
\newblock In {\em USENIX Security Symposium}, pages 327--341, 2018.

\bibitem{grinsztajn2022tree}
L{\'e}o Grinsztajn, Edouard Oyallon, and Ga{\"e}l Varoquaux.
\newblock Why do tree-based models still outperform deep learning on typical tabular data?
\newblock {\em Advances in Neural Information Processing Systems}, 35:507--520, 2022.

\bibitem{hastie2009elements}
Trevor Hastie, Robert Tibshirani, and Jerome Friedman.
\newblock {\em The Elements of Statistical Learning: Data Mining, Inference, and Prediction}.
\newblock Springer, 2001.

\bibitem{xforce}
{IBM}.
\newblock X-force threat intelligence index 2024.
\newblock \url{https://web.archive.org/web/20250529175340/https://www.ccinfo.nl/_downloads/c9f7bfb1d157ece7d96be9ed9d986d2c}, 2024.

\bibitem{wayback}
{Internet Archive}.
\newblock Wayback machine.
\newblock \url{http://web.archive.org}.

\bibitem{jacobs2021exploit}
Jay Jacobs, Sasha Romanosky, Benjamin Edwards, Idris Adjerid, and Michael Roytman.
\newblock Exploit prediction scoring system ({EPSS}).
\newblock {\em Digital Threats: Research and Practice}, 2(3):1--17, 2021.

\bibitem{jacobs2023enhancing}
Jay Jacobs, Sasha Romanosky, Octavian Suciu, Ben Edwards, and Armin Sarabi.
\newblock Enhancing vulnerability prioritization: Data-driven exploit predictions with community-driven insights.
\newblock In {\em IEEE European Symposium on Security and Privacy Workshops}, pages 194--206, 2023.

\bibitem{kotzias2018coming}
Platon Kotzias, Abbas Razaghpanah, Johanna Amann, Kenneth~G Paterson, Narseo Vallina-Rodriguez, and Juan Caballero.
\newblock Coming of age: A longitudinal study of {TLS} deployment.
\newblock In {\em Internet Measurement Conference}, pages 415--428. ACM, 2018.

\bibitem{kumar2018tracking}
Deepak Kumar, Zhengping Wang, Matthew Hyder, Joseph Dickinson, Gabrielle Beck, David Adrian, Joshua Mason, Zakir Durumeric, J~Alex Halderman, and Michael Bailey.
\newblock Tracking certificate misissuance in the wild.
\newblock In {\em IEEE Symposium on Security and Privacy}, pages 785--798, 2018.

\bibitem{kure2022asset}
Halima~Ibrahim Kure, Shareeful Islam, Mustansar Ghazanfar, Asad Raza, and Maruf Pasha.
\newblock Asset criticality and risk prediction for an effective cybersecurity risk management of cyber-physical system.
\newblock {\em Neural Computing and Applications}, 34(1):493--514, 2022.

\bibitem{lee2025using}
Jung~Youn Lee, Joonhyuk Yang, and Eric~T Anderson.
\newblock Using grocery data for credit decisions.
\newblock {\em Management Science}, 71(4):2753--2777, 2025.

\bibitem{liu2015cloudy}
Yang Liu, Armin Sarabi, Jing Zhang, Parinaz Naghizadeh, Manish Karir, Michael Bailey, and Mingyan Liu.
\newblock Cloudy with a chance of breach: Forecasting cyber security incidents.
\newblock In {\em USENIX Security Symposium}, pages 1009--1024, 2015.

\bibitem{lundberg2017unified}
Scott~M Lundberg and Su-In Lee.
\newblock A unified approach to interpreting model predictions.
\newblock {\em Advances in Neural Information Processing Systems}, 30, 2017.

\bibitem{niculescu2005predicting}
Alexandru Niculescu-Mizil and Rich Caruana.
\newblock Predicting good probabilities with supervised learning.
\newblock In {\em International Conference on Machine learning}, pages 625--632, 2005.

\bibitem{pan2017dark}
Xiaorui Pan, Xueqiang Wang, Yue Duan, XiaoFeng Wang, and Heng Yin.
\newblock Dark hazard: Large-scale discovery of unknown hidden sensitive operations in {Android} apps.
\newblock In {\em Network and Distributed System Security Symposium}, pages 1--15. Internet Society, 2017.

\bibitem{pochat2018tranco}
Victor~Le Pochat, Tom Van~Goethem, Samaneh Tajalizadehkhoob, Maciej Korczy{\'n}ski, and Wouter Joosen.
\newblock Tranco: A research-oriented top sites ranking hardened against manipulation.
\newblock {\em arXiv preprint arXiv:1806.01156}, 2018.

\bibitem{rege2023free}
Aunshul Rege and Rachel Bleiman.
\newblock A free and community-driven critical infrastructure ransomware dataset.
\newblock In {\em International Conference on Cybersecurity, Situational Awareness and Social Media}, pages 25--37. Springer, 2023.

\bibitem{rhode2018early}
Matilda Rhode, Pete Burnap, and Kevin Jones.
\newblock Early-stage malware prediction using recurrent neural networks.
\newblock {\em Computers \& Security}, 77:578--594, 2018.

\bibitem{sabottke2015vulnerability}
Carl Sabottke, Octavian Suciu, and Tudor Dumitraș.
\newblock Vulnerability disclosure in the age of social media: Exploiting {Twitter} for predicting real-world exploits.
\newblock In {\em USENIX Security Symposium}, pages 1041--1056, 2015.

\bibitem{sarabi2018characterizing}
Armin Sarabi and Mingyan Liu.
\newblock Characterizing the {Internet} host population using deep learning: A universal and lightweight numerical embedding.
\newblock In {\em Internet Measurement Conference}, pages 133--146. ACM, 2018.

\bibitem{sarabi2016risky}
Armin Sarabi, Parinaz Naghizadeh, Yang Liu, and Mingyan Liu.
\newblock Risky business: Fine-grained data breach prediction using business profiles.
\newblock {\em Journal of Cybersecurity}, 2(1):15--28, 2016.

\bibitem{sarabi2023llm}
Armin Sarabi, Tongxin Yin, and Mingyan Liu.
\newblock An {LLM}-based framework for fingerprinting {Internet}-connected devices.
\newblock In {\em Internet Measurement Conference}, pages 478--484. ACM, 2023.

\bibitem{scheitle2018first}
Quirin Scheitle, Taejoong Chung, Jens Hiller, Oliver Gasser, Johannes Naab, Roland van Rijswijk-Deij, Oliver Hohlfeld, Ralph Holz, Dave Choffnes, Alan Mislove, and Georg Carie.
\newblock A first look at certification authority authorization ({CAA}).
\newblock {\em ACM SIGCOMM Computer Communication Review}, 48(2):10--23, 2018.

\bibitem{shen2018tiresias}
Yun Shen, Enrico Mariconti, Pierre~Antoine Vervier, and Gianluca Stringhini.
\newblock Tiresias: Predicting security events through deep learning.
\newblock In {\em ACM SIGSAC Conference on Computer and Communications Security}, pages 592--605, 2018.

\bibitem{soska2014automatically}
Kyle Soska and Nicolas Christin.
\newblock Automatically detecting vulnerable websites before they turn malicious.
\newblock In {\em USENIX Security Symposium}, pages 625--640, 2014.

\bibitem{spamhaus}
The {Spamhaus} project.
\newblock \url{https://web.archive.org/web/20250519124317/https://www.spamhaus.org}.

\bibitem{statescoop}
{StateScoop}.
\newblock Ransomware attacks map.
\newblock \url{https://web.archive.org/web/20250320073527/https://statescoop.com/ransomware-map}.

\bibitem{suciu2022expected}
Octavian Suciu, Connor Nelson, Zhuoer Lyu, Tiffany Bao, and Tudor Dumitraș.
\newblock Expected exploitability: Predicting the development of functional vulnerability exploits.
\newblock In {\em USENIX Security Symposium}, pages 377--394, 2022.

\bibitem{tavabi2018darkembed}
Nazgol Tavabi, Palash Goyal, Mohammed Almukaynizi, Paulo Shakarian, and Kristina Lerman.
\newblock {DarkEmbed}: Exploit prediction with neural language models.
\newblock In {\em AAAI Conference on Artificial Intelligence}, volume~32, 2018.

\bibitem{naics}
{United States Census Bureau}.
\newblock {NAICS} codes \& understanding industry classification systems.
\newblock \url{https://web.archive.org/web/20250506224832/https://www.census.gov/programs-surveys/economic-census/year/2022/guidance/understanding-naics.html}.

\bibitem{sme}
{U.S. Small Business Administration Office of Advoacy}.
\newblock Frequently asked questions about small business.
\newblock \url{https://web.archive.org/web/20250517030943/https://advocacy.sba.gov/2023/03/07/frequently-asked-questions-about-small-business-2023}, 2023.

\bibitem{vcdb}
{Verizon}.
\newblock The {VERIS} community database ({VCDB}).
\newblock \url{https://web.archive.org/web/20250126114847/https://verisframework.org/vcdb.html}.

\bibitem{weichselbraun2021}
Albert Weichselbraun.
\newblock Inscriptis -- {A} {Python}-based {HTML} to text conversion library optimized for knowledge extraction from the web.
\newblock {\em Journal of Open Source Software}, 6(66):3557, 2021.

\bibitem{woods2021sok}
Daniel~W Woods and Rainer B{\"o}hme.
\newblock {SoK}: Quantifying cyber risk.
\newblock In {\em IEEE Symposium on Security and Privacy}, pages 211--228. IEEE, 2021.

\end{thebibliography}

\end{document}